\documentclass[11pt]{article}

\usepackage{graphicx,amssymb}

\long\def\inst#1{\par\nobreak\kern 4pt\nobreak
    {\it #1}\par\vskip 10pt plus 3pt minus 3pt}

\def\missPT{{P\mkern-11mu\slash}_T}

\begin{document}
{\pagestyle{empty}

\begin{flushright}
May 23, 2007 \\
\end{flushright}

\par\vskip 3cm

\begin{center}
{\Large \bf \boldmath
Catfish: A Monte Carlo simulator for black holes at the LHC} \\
\vspace*{1.0cm} 
{M.\ Cavagli\`a${}^a$, R.\ Godang${}^{a,b}$, L.\ Cremaldi${}^a$ and D.\ Summers${}^a$ \\
\vspace*{0.3cm}   
{\em ${}^{(a)}$ Department of Physics and Astronomy,
University of Mississippi\\
University, MS 38677-1848, USA\\
${}^{(b)}$ University of South Alabama, Department of Physics, Mobile AL 36688, USA}}
\end{center}
\bigskip

% Abstract
\begin{center}
\bf Abstract
\end{center}
\vspace*{1.0cm}
{We present a new Fortran Monte Carlo generator to simulate black hole events at CERN's
Large Hadron Collider. The generator interfaces to the PYTHIA Monte Carlo fragmentation code.
The physics of the BH generator includes, but not limited to, inelasticity effects,
exact field emissivities, corrections to semiclassical black hole evaporation and
gravitational energy loss at formation. These features are essential to realistically
reconstruct the detector response and test different models of black hole formation and
decay at the LHC.}
\vskip 24pt

\leftline{PACS: 04.80.Cc, 04.50.+h, 12.60.-i, 13.85.-t}
\leftline{Keywords: Extra dimensions, black holes, hadron colliders, Monte Carlo
methods}
\vfill 
\begin{center}
\vspace{0.1cm}\hrule\vspace{0.1cm}
Work supported in part by the U.S. Department of Energy contract DE-FG05-91ER40622.
\end{center}

}% End of pagestyle{empty}

\section{Introduction}
The fundamental scale of gravity may be much lower than the measured
gravitational scale \cite{Arkani-Hamed:1998rs}. In scenarios with large or warped
extra dimensions, the observed weakness of gravity is explained by assuming that
Standard Model (SM) fields are constrained to propagate on a four-dimensional
submanifold, whereas gravitons propagate in the higher-dimensional spacetime
\cite{Maartens:2003tw}. If the gravitational coupling constant is of the order of
few TeVs, super-Planckian events at CERN's LHC could lead to the formation of
subnuclear Black Holes (BHs) \cite{Banks:1999gd} and branes \cite{Ahn:2002mj} (For
reviews and further references, see Refs.\ \cite{Cavaglia:2002si, Cardoso:2005jq}).

The semiclassical limit of super-Planckian scattering suggests that the cross section for
creation of a BH or brane with radius $R$ is approximately equal to the geometrical Black
Disk (BD) cross section $\sigma_{BD}(s,n)=\pi R^2(s,n)$, where $\sqrt{s}$ is the Center of
Mass (CM) energy of the colliding quanta and $n$ is the number of extra dimensions. The
semiclassical Hawking effect \cite{Hawking:1974sw} provides a thermal decay mechanism for
BHs, thus allowing their detection. The spectrum of massive excitations in string theories
suggests that branes may also decay thermally \cite{Amati:1999fv}. Under the most favorable
circumstances, the BH event rate at the LHC should be comparable to the $t\bar t$ event
rate. 

Until now, numerical studies of observational signatures have used Monte Carlo (MC)
generators implementing the semiclassical picture outlined above. Currently, there are two
MC generators for BH production at particle colliders: TRUENOIR \cite{Dimopoulos:2001en} and
CHARYBDIS \cite{Harris:2003db}. However, recent results have modified significantly our
understanding of BH formation and evolution. It is thus timely and worthwile to examine the
observational signatures of BH events beyond the simple semiclassical picture. To this
purpose, we have developed a new Fortran MC generator for BH events at the LHC which
includes many of the accepted theoretical results in the literature. The generator, called
CATFISH (Collider grAviTational FIeld Simulator for black Holes), interfaces to the PYTHIA
MC fragmentation code \cite{Sjostrand:2006za} using the Les Houches interface
\cite{Boos:2001cv}. CATFISH allows the most accurate description of BH events at the TeV
scale up-to-date. Its flexibility permits to compare the signatures of different theoretical
models of BH production. MC generators with similar characteristics of CATFISH have already
been successfully utilized to simulate BH production in ultrahigh-energy cosmic ray air
showers \cite{Ahn:2005bi} and in lepton colliders \cite{Godang:2004bf}. Precompiled
executable versions of CATFISH (Linux and Mac OS platforms) are available at the CATFISH
website {\tt http://www.phy.olemiss.edu/GR/catfish}.
\section{Basics of BH formation and evolution}
In this section we follow Ref.\ \cite{Ahn:2005bi} and briefly review the basics
of BH formation and evolution. 
\subsection{BH formation and cross section at parton level}
Thorne's hoop conjecture \cite{hoop} states that an event horizon forms when a mass
$M$ is compacted into a region with circumference smaller than twice the
Schwarzschild radius $R(M)$ in any direction. At the LHC, this process can be
achieved by scattering two partons $ij$ with CM energy larger than $M$ and impact
parameter smaller $b$ than $R$. The BH event is described by the inelastic process
$ij\to {\rm BH}+E(X)$, where $E(X)$ denotes the collisional energy that does not fall
beyond the event horizon. Due to the gravitational nature of the process, this energy
includes mainly a bulk component of gravitational radiation, although non-SM gauge
fields and a brane component of SM fields cannot be excluded. If $E(X)$ is zero, the
hoop conjecture implies that the parton cross section for BH production is equal to
the geometrical BD cross section,
$\sigma_{ij}(s,n)=\sigma_{BD}(s_{ij},n)\theta(R(s_{ij})-b)$. If $E(X)\not =0$, the
cross section is generally smaller and depends on the impact parameter. Note that
this treatment is valid only if the BH is larger than the Compton length of the
colliding quanta. (For discussions on the effect of wave packet size on the BH
formation process, see Refs.\ \cite{Voloshin:2001vs}.) A precise calculation of the
collisional energy loss is essential to understanding BH formation.  

The hoop conjecture has been tested by different methods  \cite{Cardoso:2005jq}, the
most popular one being the Trapped-Surface (TS) approach
\cite{Yoshino:2002tx,Yoshino:2005hi,Vasilenko:2003ak}, The TS model gives a bound on the
inelasticity by modeling the incoming partons as two Aichelburg-Sexl shock waves
\cite{Aichelburg:1970dh}. The Aichelburg-Sexl wave is obtained by boosting the
Schwarzschild solution to the speed of light at fixed energy. The resulting metric
describes a plane-fronted gravitational shock wave corresponding to the
Lorentz-contracted longitudinal gravitational field. The parton scattering is simulated
by superposing two shock waves traveling in opposite directions. The union of these
shock waves defines a closed TS that allows to set a lower bound on the initial BH mass
$M_{BH}$. The collisional energy loss depends on the impact parameter and increases as
the number of spacetime dimensions increases. The BH mass monotonically decreases with
the impact parameter from a maximum of about 60-70\% of the CM energy for head-on
collisions.

The TS result is consistent within one order of magnitude with the hoop
conjecture. However, this approach neglects mass, spin, charge and finite-size
effects of the incoming partons. Size and spin effects are expected to be mostly
relevant around the Planck energy. Charge effects could dominate at higher energy.
The pointlike approximation fails for directions transversal to the motion
\cite{Kohlprath:2002yh}. Even with these assumptions, the TS model
provides only a lower bound on $M_{BH}$. Independent estimates of the
gravitational collisional energy loss are possible through alternative approaches.
The gravitational energy emission in a hard instantaneous collision can be evaluated
in the linearized limit \cite{Cardoso:2002pa}. This computation suggests that the
TS method overestimates the gravitational energy emitted in the process.
For head-on collisions, the instantaneous method predicts that the gravitational
energy loss is only about 10\% of the CM energy. This result is in agreement with
perturbative calculations modeling the parton-parton collision as a plunge of a
relativistic test particle into a BH with mass equal to the CM energy
\cite{Berti:2003si}.  

In conclusion, a conservative estimate of gravitational loss in relativistic
scattering at parton level gives a BH mass ranging between 60\% and 100\% of
the CM energy. The TS result and the BD result can be
considered as the lower and upper bounds on $M_{BH}$, respectively.
\subsection{Cross section at nucleon level}
The total cross section for a super-Planckian BH event involving two nucleons
is obtained by integrating the parton cross section over the Parton
Distribution Functions (PDFs). If the BH mass depends on the impact parameter, the
generally accepted formula for the total cross section of the pp process is
\begin{equation}
\sigma_{pp \to BH}(s,n) = \sum_{ij}\int_0^1 2z dz
\int_{x_m}^1dx\,\int_x^1 {dx'\over
x'}\,f_i(x',Q)f_j(x/x',Q)\,F\,\sigma_{BD}(xs,n)\,,
\label{totcross}
\end{equation}
where $f_i(\cdot,Q)$ are the PDFs with four-momentum transfer squared $Q$
\cite{Brock:1993sz,pdg2006} and $z$ is the impact parameter normalized to its maximum value.
The cutoff at small $x$ is $x_m=M_{min}^2/(sy^2(z))$, where $y(z)$ and $M_{min}$ are the
fraction of CM energy trapped into the BH and the minimum-allowed mass of the gravitational
object, respectively. $F$ is a form factor. The total cross section for the BD model is
obtained by setting $F=1$ and $y^2(z)=1$. 

Different sets of PDFs are defined in the literature. The PDFs are not known at
energies above the TeV and for values of momentum transfer expected in BH formation.
Equation (\ref{totcross}) is usually calculated by imposing a cut-off at these
values. The PDFs also suffer from uncertainties at any momentum transfer ($\sim$
10\%) \cite{Ahn:2003qn} and from the ambiguity in the definition of $Q$
\cite{Emparan:2001kf}. The momentum transfer is usually set to be $M_{BH}$ or the
Schwarzschild radius inverse. The uncertainty due to this ambiguity is about $\sim
10 - 20$\%.

The form factor and the amount of trapped energy depend in principle on energy,
gravitational scale, geometry and physical properties of the spacetime. The
TS method gives numerical values of order unity for these quantities.
(See Refs.\ \cite{Yoshino:2002tx, Yoshino:2005hi} and discussion above). However,
these results depend on the way the TS is identified. Other models
\cite{Yoshino:2005ps} give values which are more or less consistent with the
TS method. It is common practice in the literature to either choose
the TS result or the simple BD model. 

The lower cutoff on the fraction of the nucleon momentum carried by the partons
is set by the minimum-allowed formation mass of the gravitational object,
$M_{min}$. This threshold is usually considered to be roughly equal to the
minimum mass for which the semiclassical description of the BH is valid.
However, this argument is based on Hawking's semiclassical theory and may not
be valid at energies equal to few times the Planck mass. For example, the
existence of a minimum spacetime length $l_{m}$ implies the lower bound on the
BH mass \cite{Cavaglia:2003qk,Cavaglia:2004jw}:
\begin{equation}
M_{ml}= \frac{n+2}{8\Gamma\left(\frac{n+3}{2}\right)}
\left(\sqrt{\pi}\,l_{m}M_\star/2\right)^{n+1} \, M_{\star}\,,
\label{minmass}
\end{equation}
where $M_\star$ is the fundamental Planck mass. BHs with mass less than $M_{ml}$
do not exist, since their horizon radius would fall below the minimum-allowed
length. Note that $M_{ml}$ grows as a power of $l_m^{n+1}$ at fixed $M_\star$.
Therefore, $M_{min}$ may be much larger than $M_\star$ for higher-dimensional
spacetimes.
\subsection{BH evolution}
It is believed that the decay of microscopic BHs happens in four distinct
stages: I. radiation of excess multipole moments (balding phase); II.
spin-down; III. Hawking evaporation; IV. final explosion or formation of a BH
remnant. 

Although some progress has been made, a quantitative description of the balding phase
and the spin-down phase is not fully known. For example, the emission of radiation
from a $(n+4)$-dimensional rotating BH on the brane is not known for spin-2 fields
\cite{Duffy:2005ns}. Due to these limitations, balding phase and spin-down phase
effects are not implemented in the current version of CATFISH. It should be stressed,
however, that balding and spin-down effects could play an important role in BH
phenomenology at the LHC.

Many papers have been devoted to the investigation of BH evaporation in higher
dimensions \cite{Kanti:2002nr}, leading to a better understanding of the Hawking
phase. Field emissivities for all SM fields have recently been calculated
\cite{Cardoso:2005vb}. For non rotating BHs and the minimal $SU(3)\times SU(2)\times
U(1)$ SM, most of the BH mass is radiated as SM quanta on the brane, although the
gravitational emission in the bulk cannot be neglected for high $n$. Two points
should be stressed \cite{Cavaglia:2003hg}: (i) it is not clear what is the effect of
rotation on BH emissivities; (ii) the field content at trans-Planckian energies is
not known. Onset of supersymmetry, for example, could lead to other evaporation
channels and large emission of undetectable non-SM quanta during the decay phase
even in absence of rotation \cite{Chamblin:2004zg}.

Quantum gravitational effects and BH recoil \cite{Frolov:2002as} could also
affect the emission of visible quanta on the brane. Examples of quantum
gravitational effects are quantum thermal fluctuations and corrections to the
Hawking thermodynamics due to the existence of a minimum length
\cite{Cavaglia:2004jw}. The existence of a minimum scale of the order of the
Planck length \cite{Garay:1994en} is a common consequence of most (if not all)
theories of quantum gravity such as string theory, non-commutative geometry,
canonical quantum gravity, etc. The presence of a cutoff at the Planck scale
leads to a modification of the uncertainty principle. Since the Hawking
thermodynamical quantities can be derived by applying the uncertainty principle
to the BH, the existence of a minimum length leads to corrections in the
thermodynamical quantities \cite{Cavaglia:2003qk,Cavaglia:2004jw}. 

At the end of the Hawking phase, the BH is expected to either non-thermally decay in a
number $n_p$ of hard quanta or leave a remnant. In either case, the lack of a theory of
quantum gravity does not allow more than a phenomenological treatment. The final decay is
usually described by setting a cutoff on the BH mass of the order of the Planck mass,
$Q_{min}\sim M_\star$, and equally distributing the energy $Q_{min}$ to $n_p$ quanta. Since
the decay is non-thermal, and in absence of any guidance from a theory of quantum gravity,
the quanta are democratically chosen among the SM Degrees of Freedom (DoFs). Note that
$Q_{min}$ does not necessarily coincide with $M_{min}$. The former gives the threshold for
the onset of quantum gravity effects in the decay phase, whereas the latter gives the
minimum-allowed mass of the classical BH in the formation process. From the above
definitions, it follows $M_{min}\geq Q_{min}$. The existence of a minimum length gives a
natural means to set $Q_{min}$. In that case, the modified thermodynamical quantities
determine the endpoint of Hawking evaporation when the BH mass reaches $M_{ml}$. This mass
can be identified with the mass of the BH remnant.
\section{BH generator}
In this section we list the main characteristics of the CATFISH generator. The
physics of BH formation and decay is determined by the following set of external
parameters and switches in the MC code:\\

\begin{tabular}{l}
Fundamental Planck scale ($M_\star$)\\
Number of large extra dimensions ($n$)\\
Gravitational loss at formation \\
Gravitational loss model \\
Minimum BH mass at formation ($M_{min}$)\\
Quantum BH mass threshold at evaporation ($Q_{min}$)\\
Number of quanta at the end of BH decay ($n_p$)\\
Momentum transfer model in parton collision\\
Conservation of electromagnetic (EM) charge\\
Minimum spacetime length ($\alpha$)\\
\end{tabular}\\

\noindent
These parameters are briefly explained below.  
\subsection{BH formation and parton cross section}
The MC generator does not require any lower or upper bound on the Planck mass
$M_\star$. However, experimental constraints exclude values of $M_\star\lesssim 1$ TeV
\cite{collider,astro} and BHs do not form at the LHC if $M_\star > 14$ TeV. Models
with one or two flat large extra dimensions are excluded experimentally
\cite{collider,astro}. Most of the theoretical models are limited to $n\leq 7$.
Therefore, the allowed number of extra dimensions $n$ ranges from 3 to 7. (Warped
scenarios such as the Randall-Sundrum models \cite{Randall:1999vf} with a single extra
dimension are experimentally viable. However, the extra dimension is warped. Since
most of the results in the literature concerning black holes at colliders have been
derived for a flat extra-dimensional scenario, we choose not to allow $n=1$ to mimic
BH production in warped models.)

CATFISH includes three models for BH formation and cross section: BD, Yoshino-Nambu (YN) TS
model \cite{Yoshino:2002tx}, and Yoshino-Rychkov (YR) improved TS model
\cite{Yoshino:2005hi}. The minimum BH mass $M_{min}$ is set in units of $M_\star$ or, if a
minimum length is present, $M_{ml}$: $X_{min}=M_{min}\geq{\rm Max}(M_\star,M_{ml})$. This
parameter is always larger than one. 
\subsection{Total and differential cross section}
The distribution of the initial BH masses is sampled from the differential cross
section $d\sigma/dM_{BH}$. CATFISH uses the (stable) cteq5 PDF distribution
\cite{Brock:1993sz,Lai:1999wy}. The use of different PDF distributions should not
significantly affect the total and differential cross sections. Therefore, different
PDF distributions are not implemented in the code. The uncertainty due to the choice
of the momentum transfer is generally larger. A logical switch allows a choice
between $M_{BH}$ or inverse Schwarzschild radius, as the definition of momentum transfer.
The part of CM energy of the pp collision which is not trapped or lost in
gravitational radiation at formation forms the beam remnant, which is hadronized by PYTHIA.
\subsection{BH evaporation}
Due to the lack of results for the balding and spin-down phases described
above, energy losses in these stages are assumed to be either negligible or
included in the energy loss during formation. Since the TS model likely
overestimates the actual energy loss, this is a reasonable assumption. 
However, we stressed above that balding and spin-down effects could
significantly affect the event signatures. We plan to include balding and
spin-down effects in updated versions of the code, as soon as theoretical
results become available.

A similar approach is used in the Hawking phase, where the MC uses only the
emissivities of non-rotating spherically-symmetric BHs \cite{Cardoso:2005vb}.
(Emissivities for rotating BHs are not fully known.) This is a reasonable
assumption, given that the BH is expected to be bald and spinless by the time the
evaporation phase begins. Moreover, intrinsic uncertainties in event reconstruction
should hide at least some of the differences between rotating and non-rotating field
emissivities. The particle content at trans-Planckian energy is assumed to be the
minimal $SU(3)\times SU(2)\times U(1)$ SM with three families and a single Higgs
boson on a thin brane. For black holes with mass $\sim$ few TeV the Hawking
temperature is generally above $100$ GeV. Therefore, all SM DoFs can be considered
massless. (Considering massive gauge bosons does not affect the conclusions
significantly.) The spin-0, -1/2 and -1 DoFs on the brane are 1 (Higgs field), 90
(quarks + charged leptons + neutrinos) and 27 (gauge bosons), respectively. The
longitudinal DoFs of the weak bosons are included in the counting. The DoFs $c_i$
and the relative emissivities $\Gamma_{{\cal P}_i}$ and $\Gamma_{{\cal R}_i}$
\cite{Cardoso:2005vb} are given in Table~I -- III, respectively. In the notations of
Ref.~\cite{Cardoso:2005vb} the total decay multiplicity is \cite{Cavaglia:2003hg}
\begin{equation}
N=\frac{(n+1)S}{4\pi}\,\frac{\sum_i c_i{\cal R}_i\Gamma_{{\cal R}_i}}
{\sum_j c_j{\cal P}_j\Gamma_{{\cal P}_j}}\,,
\label{multin}
\end{equation}
where $S$ is the initial entropy of the BH and the emissivity normalizations
for spin-$s$ fields are:
\begin{equation}
{\cal P}_s=\left\{
\begin{array}{ll}
2.9\times 10^{-4}&s=0\\
1.6\times 10^{-4}&s=1/2\\
6.7\times 10^{-5}&s=1\\
1.5\times 10^{-5}&s=2
\end{array}\,,
\right.\qquad
{\cal R}_s=\left\{
\begin{array}{ll}
1.4\times 10^{-3}&s=0\\
4.8\times 10^{-4}&s=1/2\\
1.5\times 10^{-4}&s=1\\
2.2\times 10^{-5}&s=2
\end{array}\,.
\right.
\end{equation}
The decay multiplicities per species $N_i$ are
\begin{equation}
N_i=N\,\frac{c_i{\cal R}_i\Gamma_{{\cal R}_i}}
{\sum_j c_j{\cal R}_j\Gamma_{{\cal R}_j}}\,.
\label{multii}
\end{equation}
The presence of a minimum length affects the evaporation phase. CATFISH uses the
dimensionless parameter $\alpha=l_m M_\star/2$ to determine the minimum length. If
there is no minimum length, i.e.\ $\alpha=0$, the MC evaporates the BH according to
the Hawking theory (with varying temperature). Alternatively, the BH evolution
proceeds according to the modified thermodynamics of Ref.\
\cite{Cavaglia:2003qk,Cavaglia:2004jw}. In both cases the evaporation ends when the BH
reaches the mass $Q_{min}$. This is set in units of $M_\star$ ($M_{ml}$) if the
minimum length is zero (nonzero). Note that the BH minimum formation mass $M_{min}$
and the endpoint of Hawking evaporation $Q_{min}$ are independent parameters. 
\begin{table}[ht]
\begin{center}
\begin{tabular}{|l||c|}
\hline
& $c_i$ \\
\hline
 ~Quarks & 72 \\
 ~Charged leptons~ &  12\\
 ~Neutrinos &  6 \\
 ~Photon &  2 \\
 ~EW bosons &   9 \\
 ~Gluons &  16 \\
 ~Higgs &  1 \\
 ~Graviton & 1 \\
\hline
\end{tabular}
\caption{DoFs $c_i$ for the SM fields on a thin brane. The graviton is assumed to
propagate in all $(n+4)$ dimensions. Following Ref.~\cite{Cardoso:2005vb}, the
$(n+4)(n+1)/2$ graviton helicities are included in the emissivities (see Table
\ref{tab:totalpower3} and \ref{tab:emission rates} below). Therefore, the graviton
counts as one DoF. Longitudinal DoFs are included in the EW boson counting.}
\end{center}
\begin{center}
\begin{tabular}{|l||ccccc|}
\hline
&n=3&n=4&n=5&n=6&n=7\\
\hline
~Higgs         &1   &1   &1  &1&1\\ 
~Fermions      &0.89    &0.87  &0.85  &0.84 &0.82\\ 
~Gauge Bosons~ &1.0  &1.04  &1.06  &1.06&1.07\\
~Gravitons     &2.7   &4.8   &8.8   &17.7 &34.7\\ 
\hline
\end{tabular}
\caption{\label{tab:totalpower3} Fraction of radiated power per DoF and
species $i$, $\Gamma_{{\cal P}_i}$, normalized to the Higgs field. The
graviton values include all the helicity states. (From Ref.\ \cite{Cardoso:2005vb}.)}
\end{center}
\begin{center}
\begin{tabular}{|l||ccccc|}
\hline
&n=3&n=4&n=5&n=6&n=7\\
\hline
~Higgs         &1   &1   &1  &1&1\\
~Fermions      &0.78    &0.76  &0.74  &0.73&0.71\\
~Gauge Bosons~ &0.83  &0.91  &0.96  &0.99&1.01\\
~Graviton      &0.91   &1.9   &2.5 &5.1  &7.6\\
\hline
\end{tabular}
\caption{\label{tab:emission rates} Fraction of emission rates per DoF and
species $i$, $\Gamma_{{\cal R}_i}$, normalized to the Higgs field. The
graviton values include all the helicity states. (From Ref.~\cite{Cardoso:2005vb}.)}
\end{center}
\end{table}

Four-momentum is conserved at each step in the evaporation process by taking
into account the recoil of the BH on the brane due to the emission of the
Hawking quanta. The initial energy of the BH is distributed democratically
among all the Hawking quanta with a random smearing of $\pm 10$\%. This
smearing factor is introduced on a purely phenomenological basis to take into
account quantum uncertainties in the emission of each quantum.
\subsection{BH final decay}
The MC code allows for two different choices of final BH decay: Final explosion
in a number $n_p$ of quanta or BH remnant. If $n_p=0$, the BH settles down to a
remnant with mass $Q_{min}$. If $n_p$ = 2\dots 18, the BH decays in a number
$n_p$ of quanta by a $n$-body process with total CM energy equal to
$Q_{min}$. 

CATFISH allows conservation of color and EM charges. Color charge is always
conserved. A logical switch controls conservation of EM charge in the decay process
(Hawking evaporation + final decay). The purpose of this switch is to allow for the
existence of a charged or neutral BH remnant.

If the EM charge switch is set to {\tt FALSE}, there is no constraint on the total
charge of the emitted quanta $Q_E$. If $n_p=0$, physical charge conservation implies
the relation $Q_E+Q_R+Q_B=2e$, where $Q_E$ is the total charge of the Hawking quanta,
$Q_R$ is the charge of the BH remnant and $Q_B$ is the charge of the beam remnant. In
that case, the BH remnant can be either neutral or charged, depending on the event.
The choice $n_p\not = 0$ and no charge conservation ({\tt FALSE}) is unphysical and
should be avoided.

If the EM charge switch is set to {\tt TRUE}, the absolute value of the total
charge of the emitted quanta is $|Q_E|\leq 4e/3$, i.e.\ the maximum possible
total charge of the scattering partons. In that case, the excess charge
$2e-Q_E$ is assigned to the beam remnant and, if $n_p=0$, the BH remnant is
considered neutral. This is justified from the fact that local charges should
have been shed earlier in the evaporation process. (See, however, Ref.\
\cite{Koch:2005ks} for a different viewpoint.) It should be stressed that the
collider phenomenology of a charged remnant is not known and it is not clear
how to track it in a detector in a meaningful way.
\subsection{Event simulation}
The steps to simulate a BH event are:
\begin{enumerate}
\item Two proton beams of energy $7+7$ TeV are injected in the Monte Carlo (CM frame).
\item The cross section for the process is computed.
\item The initial black hole mass is sampled from the differential cross section (see Fig.\
1). 
\item The black hole is decayed through the Hawking mechanism and final hard event (or black
hole remnant).
\item The unstable quanta from the black hole and beam remnant are hadronized or decayed
instantaneously by PYTHIA. Initial- and final-state radiation are included in
PYTHIA's output.
\end{enumerate}
\section{Analysis of BH events}
Signatures of BH events at the LHC have been investigated in a number of papers 
\cite{Stocker:2006we,Harris:2004xt,Alberghi:2006qr,Tanaka:2004xb,Lonnblad:2005ah,Nayak:2006vf}
using the TRUENOIR \cite{Dimopoulos:2001en} or CHARYBDIS \cite{Harris:2003db}
generators. In this section we present some results for CATFISH. We focus on a purely
statistical analysis of variables which allow an easy comparison with previous results
obtained with the CHARYBDIS generator. A more refined analysis of other detector
response-dependent signatures such as back-to-back di-jet suppression, di-lepton
events ($\mu^+\mu^-$, $\mu^+e^-$, $\mu^+e^+$, \dots) will be presented in a
future publication \cite{inprep}. 
\subsection{Visible energy and visible/missing transverse momentum \label{vismiss}} 
Missing transverse momentum ($\missPT$) and visible transverse momentum of leptons
and hadrons are important signatures of BH production in particle colliders.  Figure
\ref{figure:2} shows the simulation output for 10,000 events at the LHC
with the following parameters (benchmark):
$$n=6\,,\qquad M_{min}=Q_{min}=M_\star\,,\qquad n_p=4\,,\qquad \alpha=0\,,$$
BD cross section and conservation of EM charge. The momentum transfer is set to be equal to
the Schwarzschild radius inverse. Particles in the beam pipe and in the inital-radiation
phase has been removed by imposing $P_T$ cuts of $5$ GeV and $15$ GeV on leptons ($e,\mu$)
and photons + hadrons ($\gamma,h$), respectively. (These choices of cuts and momentum
transfer apply to all simulations throughout the paper.) The plots show the total visible
energy distribution, $\missPT$ and the visible transverse momentum of leptons ($e,\mu$) and
photons + jets ($\gamma,h$) with varying fundamental scale $M_\star=1\dots 3$ TeV.

The plots in Fig.\ \ref{figure:2} for the BD model can be used to compare CATFISH with
previous BH generators. For example, these results are in good agreement with results
obtained with CHARYBDIS \cite{Harris:2004xt}. BH events may show a large amount of transverse
momentum up to several TeV, depending on the value of the fundamental scale and the number of
extra dimensions. 

In the absence of a BH remnant and for the BD model, the missing transverse momentum is due
to undetectable quanta (gravitons + neutrinos) during the evaporation phase. Detectable
quanta are originated in the Hawking and final decay phase with an upper bound to their
multiplicity given by $N+n_p$, where $N$ is given in Eq.\ (\ref{multin}). The bulk of BH
events is characterized by light, low-entropy BHs. Since the graviton and neutrino channels
accounts only for a small fraction of the total multiplicity in the decay phase, only rare
high-mass events show a large amount of missing transverse momentum. A rough counting of DoFs
shows that the hadronic-to-leptonic decay ratio of a BH event should be approximately 5:1.
The prevalence of the hadronic channel on the leptonic channel is evident from the lower
panels of Fig.\ \ref{figure:2}.  

Figures \ref{figure:2} also shows the effect of the fundamental scale on visible energy and
missing and visible transverse momentum. Increasing $M_\star$ leads to more massive BHs,
i.e., higher multiplicity and harder quanta in the Hawking phase. Therefore, higher values of
$M_\star$ tend to produce larger $\missPT$. The visible transverse momenta show a similar
pattern. Observation of events with high $\missPT$ would indicate high values of $M_\star$,
independently of the details of BH formation and the number of extra dimensions. If BHs are
observed at the LHC, it is thus conceivable that $M_\star$ could be measured to a certain
degree of precision. 

Changing the number of extra dimensions affects the BH mass and the missing and visible
energy outputs. Graviton emission increases with the number of extra dimensions
\cite{Cardoso:2005vb}, leading to a decrease in visible energy for high $n$. The variation in
$\missPT$ due change in spacetime dimesionality is much less significant due to the high
degree of sphericity of BH events (upper-right panels). Effects due to the dimensionality of
spacetime are more evident for massive BHs, whereas most of the BHs produced at the LHC are
very light. Therefore, it is likely that LHC would not be able to determine the number of
extra dimensions just by statistical means. 

Figure \ref{figure:3} shows the effects of varying the minimum mass lower bound. The
distributions separate quite easily the two values of $M_{min}$. However, since $M_{min}$ is
a lower bound on the BH mass, increases in $M_{min}$ are akin to increases in $M_\star$
(compare the upper-left panels of Fig.\ \ref{figure:2} and Fig.\ \ref{figure:3}). Changes in
$M_{min}$ are also entangled with the initial graviton emission, in particular for massive
events: In the BD model, larger values of $M_{min}$ (at fixed $M_\star$) lead to more massive
BHs, and thus to higher visible transverse momenta. If the initial gravitational emission is
turned on, this increase may be balanced by a decrease due to lower multiplicity (compare
$M_{min}=1$ TeV for the BD model with $M_{min}=2$ TeV for the YR model). A measure of
$M_{min}$ on purely statistical basis might prove to be difficult at the LHC. 

Figure \ref{figure:4} displays the effects of the final BH decay (YR TS model). The
distributions show that it is virtually impossible to distinguish the $n_p=2$ model from the
$n_p=4$ model. Although quanta emissivities in the Hawking phase and the final phase differ
for the presence of greybody factors in the former, the difference is not sufficient to allow
a separation without a spectral analysis of the energy and the number of emitted quanta. The
detection of a BH remnant stands a better chance because of larger $\missPT$ and smaller
visible momentum due to its undetectability. (See also Refs.\
\cite{Koch:2005ks,Stocker:2006we}.) Note that a large fraction of events with remnant
produces very little visible output. This is due to the fact that most of the BHs are
initially so light that the Hawking phase does not take place. For higher mass events, the
energy carried by the decay products is much larger than the invisible energy carried by the
remnant. Therefore, detection of a remnant is less likely in high-mass events. 

Figure \ref{figure:5} compares a smooth spacetime and a spacetime with nonzero minimum length
equal to the fundamental Planck scale inverse. The plots show no significant statistical
differences between the two cases. The effects of a small distance cut-off becomes only
relevant when the minimum scale is very close to the threshold of complete suppression of BH
production, i.e., when the minimum allowed mass Eq.\ (\ref{minmass}) is so large that BHs
cannot form at the LHC CM energy. Therefore, observation of minimum length effects at the LHC
requires a certain degree of fine tuning in the parameter $\alpha$.
\subsection{Sphericity, thrust and Fox-Wolfram moments} 
BH events are expected to be highly spherical because of the spherical nature of Hawking
evaporation. The event shape can be quantified by means of the sphericity $S$
and aplanarity $A$ \cite{Bjorken:1969wi}, thrust and oblateness $T$ \cite{Brandt:1964sa}, and
Fox-Wolfram moment $R_1\dots R_4$ variables \cite{Fox:1978vu}. Fig.\ \ref{figure:6} shows
sphericity, aplanarity, oblateness and thrust for
$$M_\star=1~{\rm TeV}\,,\qquad n=6\,,\qquad M_{min}=Q_{min}=M_\star\,,\qquad\alpha=0\,,$$
BD and TS models and different final decay modes ($n_p=2,4$), respectively. (Rare) massive BH
events are characterized by very high sphericity and isotropy. A similar conclusion is
reached by examining the second Fox-Wolfram moment (see first panel of Fig.\ \ref{figure:7}).
Increasing $M_{min}$ makes the events even more spherical because of the higher multiplicity
in the decay phase.

Comparison between BD and TS models at fixed $n_p$ shows that the BD model leads on average
to more spherical events. This is expected because BD BHs are more massive and emit more
quanta in the Hawking phase than TS BHs. The higher sphericity of BD events is evident from
the central part of the distributions, where Hawking emission dominates the emission in the
final explosive phase, making the statistical difference between BD and TS models more clear.
Comparison between $n_p=2$ and $n_p=4$ at fixed BD or TS shows the former to be less
spherical than the latter. This effect is better displayed in the region of the plots
corresponding to light BHs, where emission in the final phase dominates over Hawking
emission. However, it should be stressed that the distinction between $n_p=2$ and $n_p=4$ at
the LHC might well prove impossible due to the presence of non-BH background. Distinction
between BD and alternative models of BH formation should be possible by selecting massive
spectacular events with high sphericity. 
\subsection{Heavy and light jet mass} 
The upper-right and the lower panels of Fig.\ \ref{figure:7} show the number of jets and
heavy and light jet mass \cite{Sjostrand:2006za} for the choice of parameters discussed
above, respectively. Note that these plots include initial- and final-state radiation jets in
addition to the jets originated by the BH decay phase. The BD model produces on average more
jets than the TS model (upper-right panel of Fig.\ \ref{figure:7}). This is also evident from
the right portions of the jet mass distributions, where the BD model is characterized by more
massive jets than the TS model at fixed $n_p$. Therefore, measurement of high jet mass allows
determination of the BH formation model independently of the shape variables. The left
portions of the jet mass distributions are sensitive to the final BH decay. Final decay in
two jets produces more heavy jets than final decay in four jets at fixed BD or TS model.
Therefore, the measurement of low jet mass may give important information on the physics of
the final BH phase.
\section{Conclusions and further developments}
The study of BH production at the TeV scale is now a few years old and entering the mature
stage. Although some of the characteristics of subatomic BH production remain obscure, many
new theoretical results have been published in the literature. A MC generator which
includes these theoretical results is needed for accurate simulations of BH events at the
forthcoming LHC. Such a generator is also important to check the stability of the overall
picture of BH production against improvements in the theory and have independent
confirmation of previous results obtained with existing generators. With this in mind, we
have developed CATFISH. The CATFISH generator implements several features of BH production
at the TeV scale which were not included in TRUENOIR and CHARYBDIS. CATFISH new physics
includes inelasticity effects during the BH formation phase
\cite{Yoshino:2002tx,Yoshino:2005hi}, exact field emissivities (albeit only for
non-rotating BHs) \cite{Cardoso:2005vb}, corrections to Hawking's semiclassical evaporation
phase \cite{Cavaglia:2003qk,Cavaglia:2004jw}, BH recoil on the brane, and different final
BH decay modes with possibility of remnant formation \cite{Koch:2005ks}. These features
allow the most accurate description of BH events at the TeV scale up-to-date. Another
important feature of CATFISH is its flexibility. CATFISH design based on independent
subroutine blocks allows easy inclusion of new theoretical results as soon as they become
available. For example, the most significant changes to the phenomenology of BH formation
in particle colliders is expected to arise from spin and charge effects. Emissivities for
rotating and/or charged BHs can be easily implemented in CATFISH if known. We are also
planning to include in future versions of the MC generator backreaction effects during the
Hawking phase (see, e.g., Ref.\ \cite{Alberghi:2006qr}), thermodynamic fluctuations
\cite{Cavaglia:2004jw}, SUSY effects \cite{Chamblin:2004zg} and photosphere and
chromosphere effects \cite{Anchordoqui:2002cp}.

The analysis of BH formation presented in the second part of this paper is limited to a few
statistical observables. This represents by no means CATFISH full potentiality. Several
other interesting signatures of BH formation in particle colliders have been investigated in
the literature (see, e.g., Refs.\
\cite{Stocker:2006we,Harris:2004xt,Tanaka:2004xb,Lonnblad:2005ah,Nayak:2006vf}). In
particular, suppression of high-energy back-to-back-correlated di-jets with energy above the
fundamental scale and di-lepton production with large transverse momentum are expected to be
two of the most interesting signatures of BH production at the LHC. Investigation of these
signatures with CATFISH is in progress \cite{inprep}. Finally, detector response and event
reconstruction are also fundamental issues to be addressed in a complete analysis of BH
events at the LHC. Further work along these lines is currently being pursued. Precompiled
executable Linux and Mac OS versions of CATFISH can be downloaded at the CATFISH website:
{\tt http://www.phy.olemiss.edu/GR/catfish}.
\section*{Acknowledgments}
The authors would like to thank Vitor Cardoso, David Cline, Greg Landsberg,
Alexander Melnitchouk, Robert Palmer, David Sinn, Janes Pinfold, Hans Wenzel and
Graham Wilson for discussions and many useful suggestions. This work was
supported in part by the U.S. Department of Energy contract  DE-FG05-91ER40622.

\newpage
\begin{figure}[ht]
\begin{center}
\centerline{\includegraphics[height=0.36\textheight]{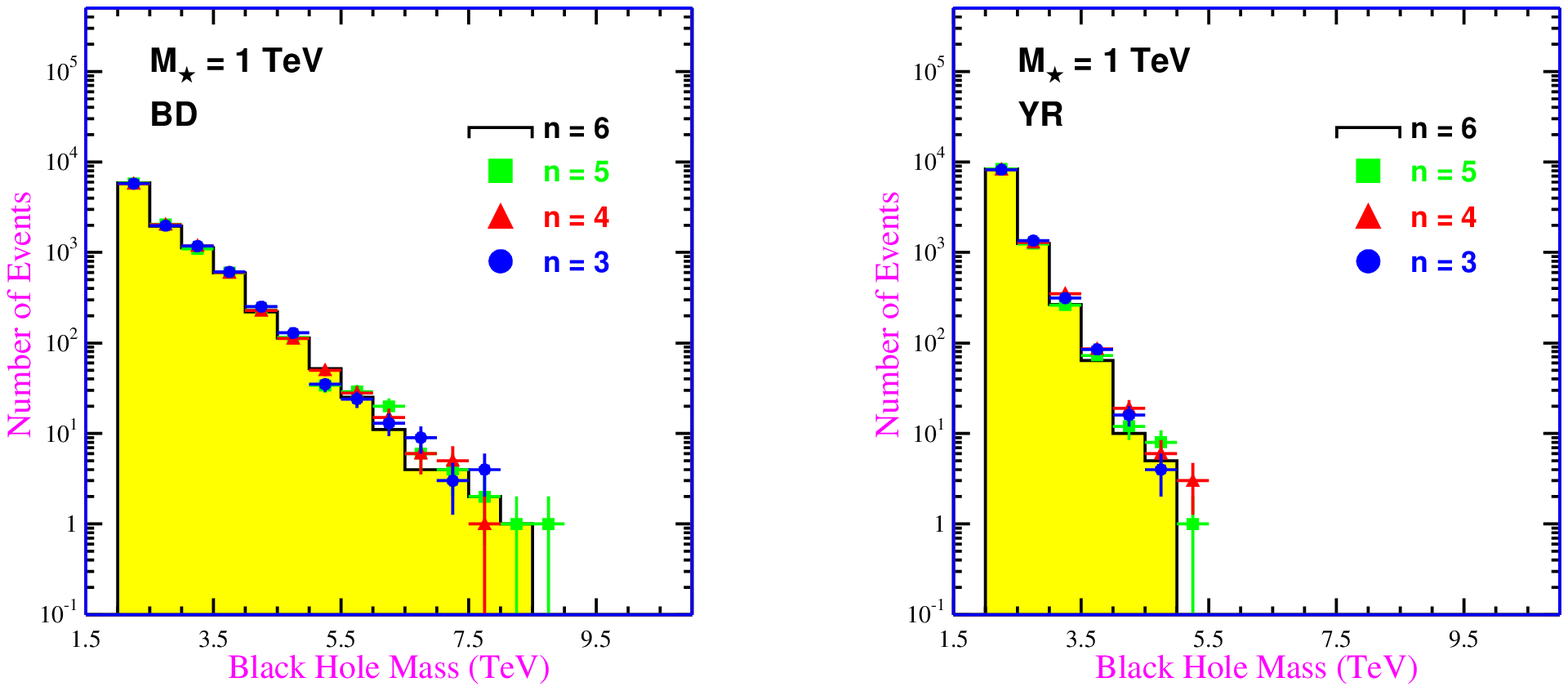}}
\vspace{-10truemm}
\caption[Black hole mass distribution for $M_\star=1$ TeV and varying
$n$.]{\label{figure:1} $M_{BH}$ distribution for the black disk model (BD) and
the Yoshino-Rychkov TS model (YR) and number of extra dimensions $n=3\dots 6$.
The fundamental Planck scale $M_\star$ is 1 TeV.}
\end{center}
\end{figure}
\newpage
\begin{figure}[ht]
\begin{center}
\centerline{\includegraphics[height=0.92\textheight]{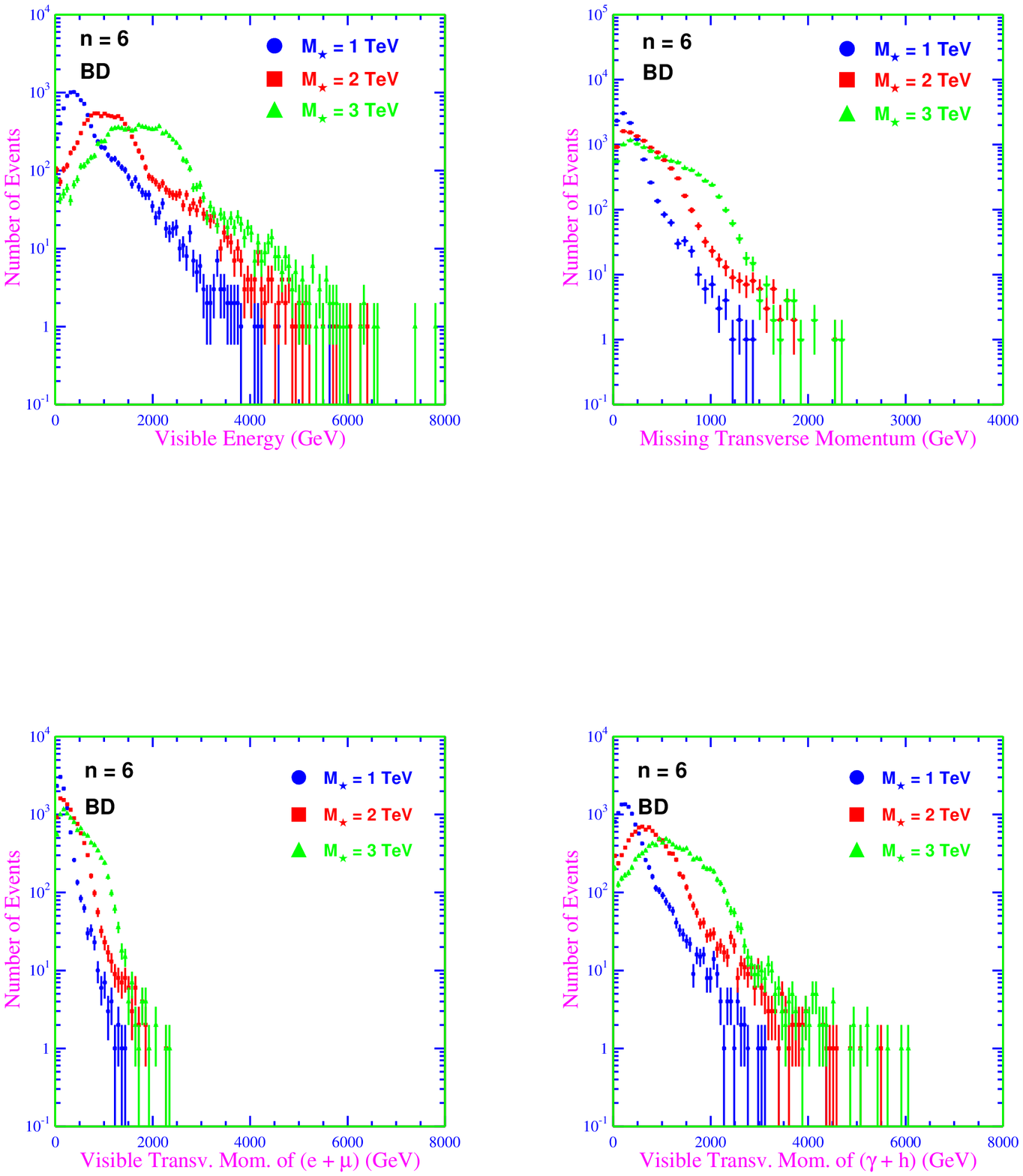}}
\vspace{-10truemm}
\caption{\label{figure:2} Visible energy, $\missPT$ and visible transverse momentum of
leptons and photons+jets (GeV) for the black disk model (BD) and fundamental Planck scale
$M_\star=1,2,3$ TeV. The number of spacetime is ten-dimensional ($n=6$). The final BH decay
is in $n_p=4$ quanta.}
\end{center}
\end{figure}
\newpage
\begin{figure}[ht]
\begin{center}
\centerline{\includegraphics[height=0.92\textheight]{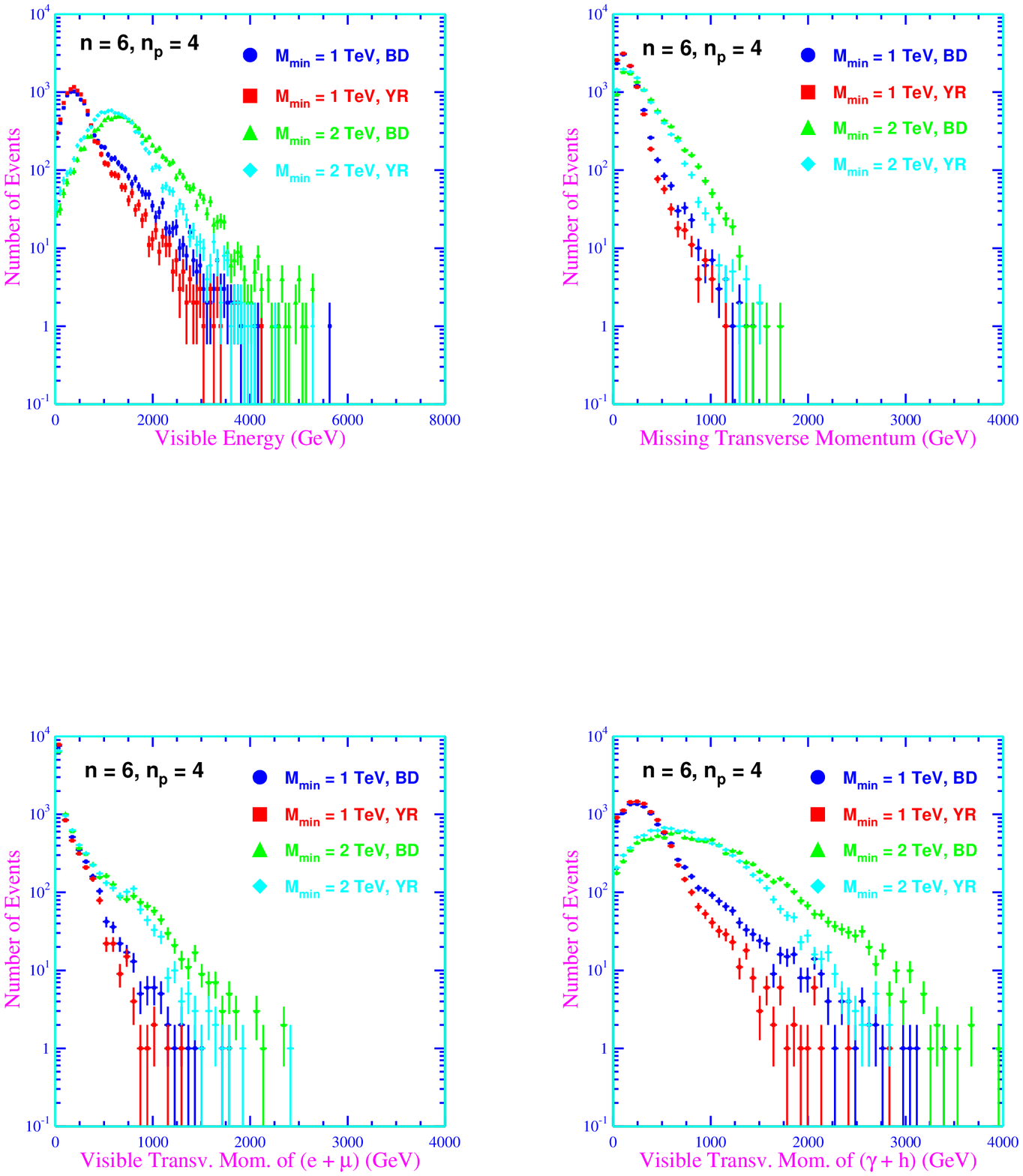}}
\vspace{-15truemm}
\caption{\label{figure:3}  Visible energy, $\missPT$ and visible transverse momentum of
leptons and photons+jets (GeV) for the black disk model (BD) and the Yoshino-Rychkov TS model
(YR). The fundamental Planck scale is $M_\star=1$ TeV. The minimum formation mass of the BH
is $M_{min}=1$ TeV or $M_{min}=2$ TeV. The final BH decay is in $n_p=4$ quanta.}
\end{center}
\end{figure}
\begin{figure}[ht]
\begin{center}
\centerline{\includegraphics[height=0.92\textheight]{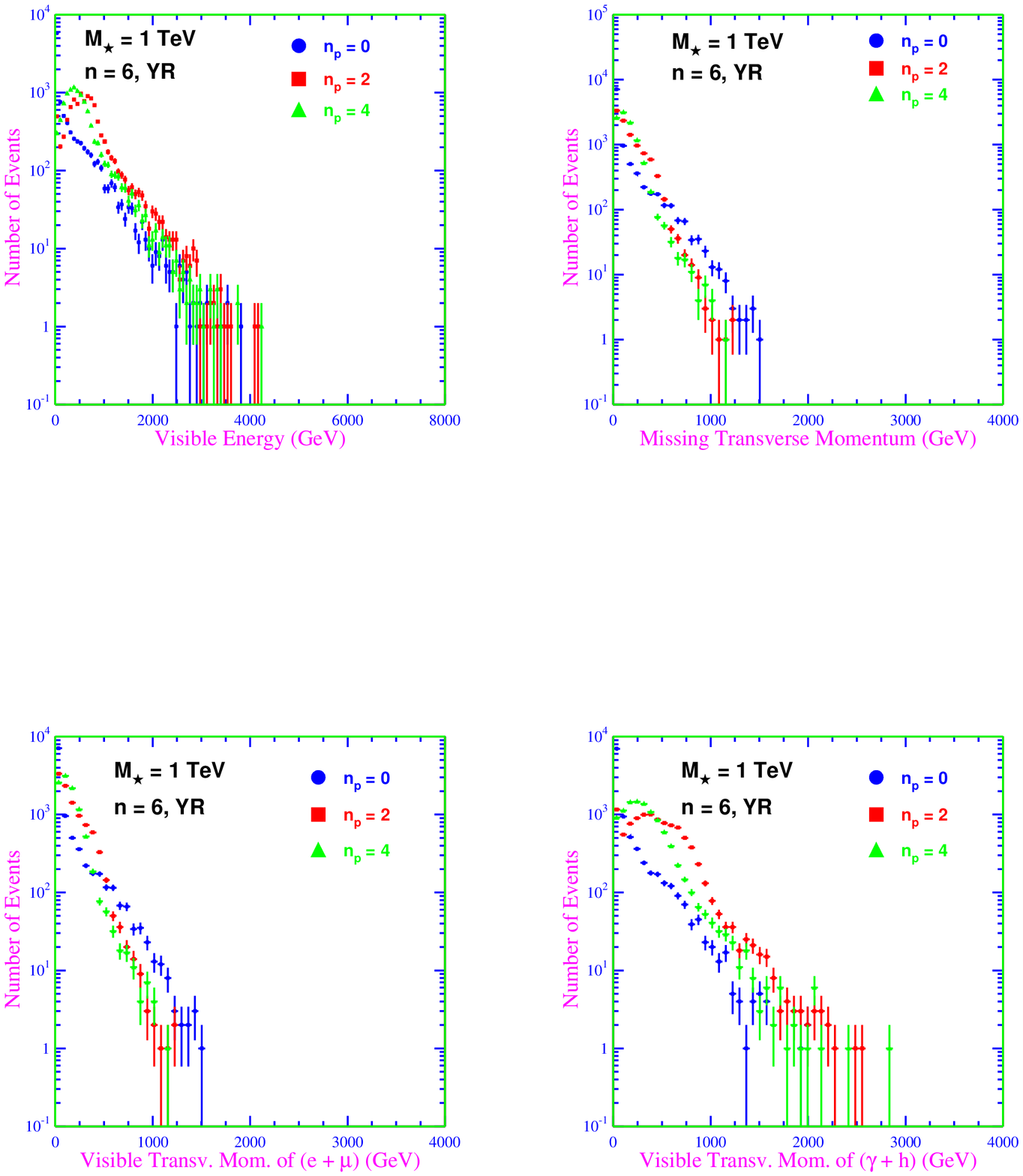}}
\vspace{-15truemm}
\caption{\label{figure:4} Visible energy, $\missPT$ and visible transverse momentum of
leptons and photons+jets (GeV) for the Yoshino-Rychkov TS model (YR) with fundamental Planck
scale $M_\star=1$ TeV and three different final decay modes: neutral remnant ($n_p=0$), two
and four quanta.}
\end{center}
\end{figure}
\begin{figure}[ht]
\begin{center}
\centerline{\includegraphics[height=0.92\textheight]{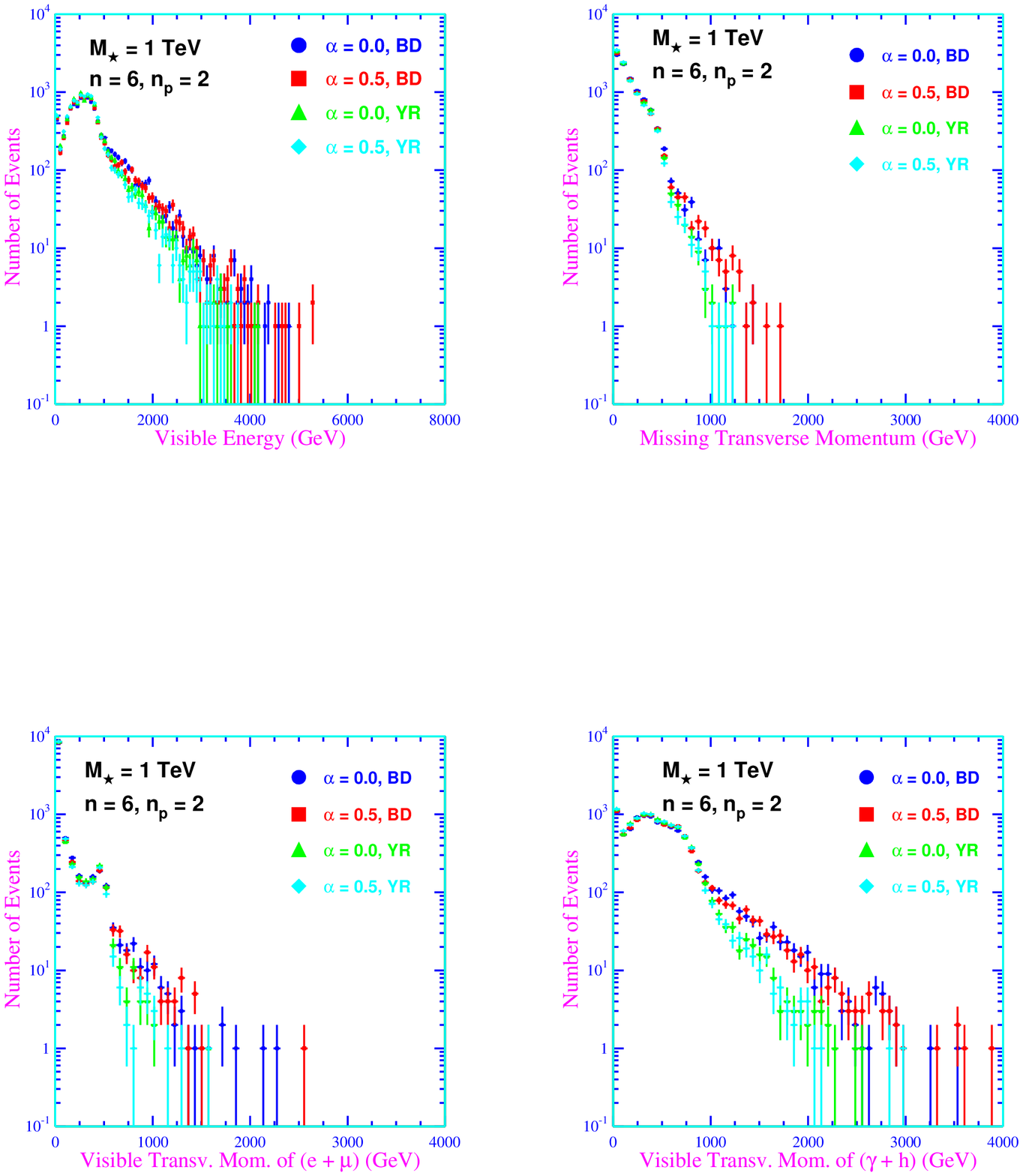}}
\vspace{-15truemm}
\caption{\label{figure:5}  Visible energy, $\missPT$ and visible transverse momentum of
leptons and photons+jets (GeV) for the black disk model (BD) and the Yoshino-Rychkov TS (YR)
model with zero ($\alpha=0$) or $M_\star^{-1}$ ($\alpha=0.5$) minimum length. The final BH
decay is in $n_p=2$ quanta.}
\end{center}
\end{figure}
\begin{figure}[ht]
\begin{center}
\centerline{\includegraphics[height=0.92\textheight]{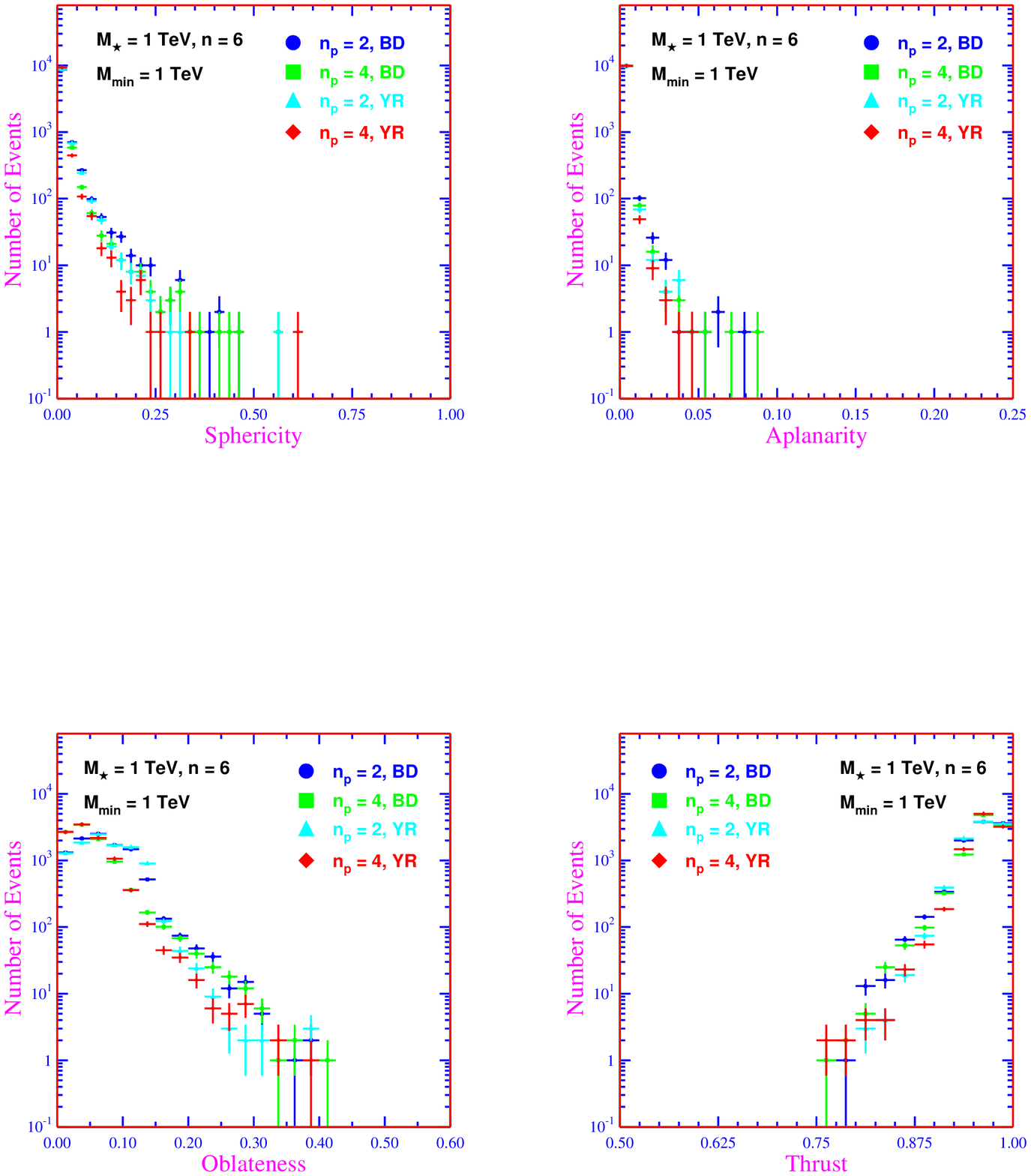}}
\null\vspace{-22truemm}
\caption{\label{figure:6}  Sphericity, aplanarity, oblateness and thrust for the black disk
model (BD) and the Yoshino-Rychkov TS model (YR). The final black hole decay is in two
$n_p=2$ or $n_p=4$ quanta.}
\end{center}
\end{figure}
\begin{figure}[ht]
\begin{center}
\centerline{\includegraphics[height=0.92\textheight]{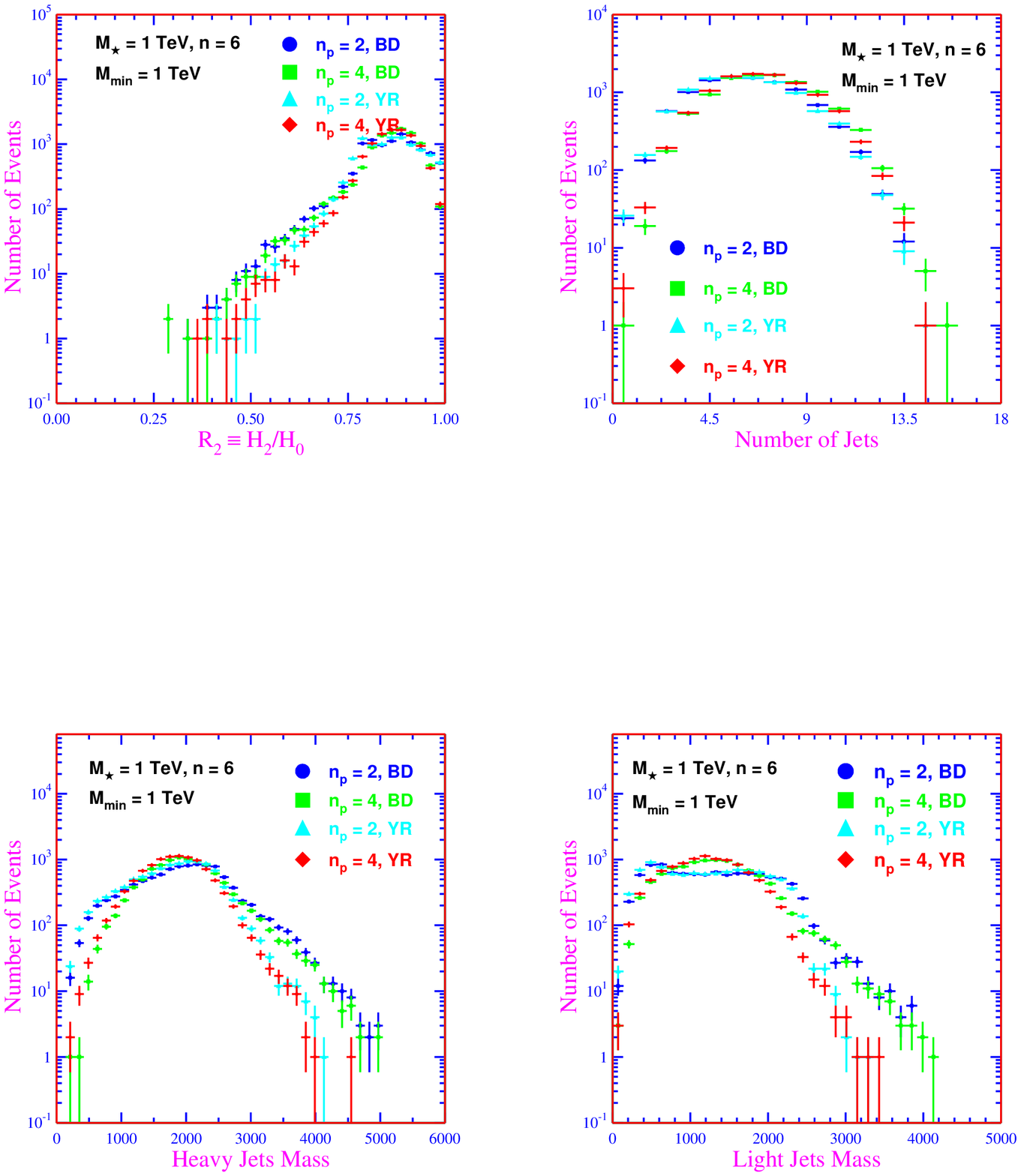}}
\null\vspace{-22truemm}
\caption{\label{figure:7} Fox-Wolfram moment $R_2$, number of jets, heavy and light jet mass
for the black disk model (BD) and the Yoshino-Rychkov TS model (YR). The final black hole
decay is in $n_p=2$ or $n_p=4$ quanta.} \end{center}
\end{figure}

\end{document}